\documentclass[conference]{template/IEEEtran}
\IEEEoverridecommandlockouts
\usepackage{cite}
\usepackage{amsmath,amssymb,amsfonts}
\usepackage{algorithmic}
\usepackage{graphicx}
\usepackage{textcomp}
\usepackage{xcolor}
\def\BibTeX{{\rm B\kern-.05em{\sc i\kern-.025em b}\kern-.08em
    T\kern-.1667em\lower.7ex\hbox{E}\kern-.125emX}}

\usepackage{booktabs}
\usepackage[hidelinks]{hyperref}
\usepackage{cleveref}
\usepackage{multirow}
\usepackage{glossaries}
\usepackage{caption}

\newacronym{ai}{AI}{artificial intelligence}
\newacronym{dear}{DEAR}{Deep Evaluation of Audio Representations}
\newacronym{dl}{DL}{deep learning}
\newacronym{drr}{DRR}{direct-to-reverberant energy ratio}
\newacronym{knn}{kNN}{$k$-nearest neighbor}
\newacronym{rt60}{RT60}{reverberation time}
\newacronym{snr}{SNR}{signal-to-noise ratio}
\newacronym{ssl}{SSL}{self-supervised learning}
\newacronym{who}{WHO}{World Health Organisation}

\begin{document}

\title{Evaluation of Deep Audio Representations\\for Hearables
\thanks{This work was supported by the Swiss Innovation Agency Innosuisse under project 100.327 IP-ICT ``DARLING''.}
}

\author{
\IEEEauthorblockN{
\hspace{2cm}
Fabian Gr\"oger\IEEEauthorrefmark{1}\IEEEauthorrefmark{3}\textsuperscript{\textsection}
}
\and
\IEEEauthorblockN{
Pascal Baumann\IEEEauthorrefmark{1}\textsuperscript{\textsection}
}
\and
\IEEEauthorblockN{
Ludovic Amruthalingam\IEEEauthorrefmark{1}
}
\and
\IEEEauthorblockN{
Laurent Simon\IEEEauthorrefmark{2}\hspace{2cm}
}
\and
\IEEEauthorblockN{
\hspace{4cm}
Ruksana Giurda\IEEEauthorrefmark{2}
}
\and
\IEEEauthorblockN{
Simone Lionetti\IEEEauthorrefmark{1}
\hspace{4cm}
}
\and
\IEEEauthorblockA{
\IEEEauthorrefmark{1}%
\textit{Lucerne University of Applied Sciences and Arts},\\
Rotkreuz, Switzerland\\
Email: \{first\}.\{last\}@hslu.ch
}
\and
\IEEEauthorblockA{
\IEEEauthorrefmark{2}%
\textit{Sonova AG},\\
St\"afa, Switzerland\\
Email: \{first\}.\{last\}@sonova.com
}
\and
\IEEEauthorblockA{
\IEEEauthorrefmark{3}%
\textit{University of Basel},\\
Basel, Switzerland\\
Email: \{first\}.\{last\}@unibas.ch
}
}

\maketitle
\frenchspacing

\begingroup
\renewcommand\thefootnote{\textsection}
\footnotetext{These authors have contributed equally.}
\endgroup

\begin{abstract}
Effectively steering hearable devices requires understanding the acoustic environment around the user.
In the computational analysis of sound scenes, foundation models have emerged as the state of the art to produce high-performance, robust, multi-purpose audio representations.
We introduce and release Deep Evaluation of Audio Representations (DEAR), the first dataset and benchmark to evaluate the efficacy of foundation models in capturing essential acoustic properties for hearables. 
The dataset includes 1,158 audio tracks, each 30 seconds long, created by spatially mixing proprietary monologues with commercial, high-quality recordings of everyday acoustic scenes.
Our benchmark encompasses eight tasks that assess the general context, speech sources, and technical acoustic properties of the audio scenes. 
Through our evaluation of four general-purpose audio representation models, we demonstrate that the BEATs model significantly surpasses its counterparts. 
This superiority underscores the advantage of models trained on diverse audio collections, confirming their applicability to a wide array of auditory tasks, including encoding the environment properties necessary for hearable steering.
The DEAR dataset and associated code are available at \url{https://dear-dataset.github.io}.
\end{abstract}


\section{Introduction}

The \gls{who} estimates that almost half a billion people suffer from disabling hearing loss~\cite{who2021world}.
This condition has been extensively proven to have a strong negative impact on social interactions and cognitive abilities~\cite{cunningham2017hearing,ha2020hearing,fisher2014impairments}. 
Hearing aids are the most common medical devices used to mitigate hearing difficulties by actively filtering and processing sound for the user. 
However, many individuals report hearing challenges even without a diagnosis of clinical hearing loss~\cite{humes2023us}. 
While these individuals are not typically prescribed hearing aids, they could benefit from features like active noise cancellation and speech enhancement, which are now available in the latest hearable devices~\cite{edwards2020emerging}.
Independently of whether they are medical devices or not, the final goal of such instruments is to help their users in challenging acoustic situations.

In order to apply the correct processing, the devices first need to analyze the surrounding environment.
This step can be challenging, especially in complex and dynamic acoustic scenes with concurrent sound sources.
Moreover, the dimensions, battery lifetime, and usability of the devices themselves set strict requirements on power consumption, available memory, possible number of operations per cycle, and maximum allowed delay.
However, the technological leap witnessed in the past years has opened the doors to solutions that used to be considered beyond reach.
Whether in a medical device, an off-the-shelf product, or a hearable, \gls{ai} can provide measurable benefits for users~\cite{goehring2016speech,zhao2018deep,healy2021effectively,andersen2021creating}. 

\Gls{dl} in particular has enabled remarkable results in audio analysis \cite{liu2022audio}, and is making an impact for hearing devices and hearables \cite{diehl2023restoring}.
A major recent trend is the development of \gls{dl} models that are not optimized to solve a single task, but learn representations that can be used for diverse purposes.
These are generally referred to as foundation models, and are typically trained with \gls{ssl} on large amounts of unlabeled data.
Thanks to their versatility, these neural networks have rapidly gained popularity, in particular for their ability to handle diverse data in their training domain and to generalize beyond their training context.
Evaluating such models requires testing their capacity to accurately encode diverse properties, which can be human-given labels or characteristics collected together with the data.
In the case of audio, this was pioneered by NOSS~\cite{shor2020towards}, LeBenchmark~\cite{evain2021lebenchmark}, and SUPERB~\cite{yang2021superb}, which focus on speech tasks, and first extended beyond speech by HARES~\cite{wang2022towards}.
The HEAR~\cite{turian2022hear} and LAPE~\cite{ghosh2022decorrelating} benchmarks followed, addressing the comprehensiveness of evaluation tasks and low-resource environments, respectively.
More recently, ARCH~\cite{la2024benchmarking} was developed to complement and extend HEAR in evaluating general-purpose audio representations.
Most closely related our work is the ACE challenge~\cite{eaton_estimation_2016}, which explicitly focuses on a general audio dataset for RT60 and DRR.

Foundation models for audio are great candidates for becoming the basic signal processing engines for hearables.
There are however still significant gaps in understanding how well these models would perform for typical hearable steering in real life.
A first major shortcoming is that none of the benchmarks mentioned above probes basic acoustic properties such as reverberation or \gls{snr}, which are crucial for signal processing in hearables.
Technical reverberation parameters such as \gls{drr} and \gls{rt60} plus \gls{snr} are known to heavily affect binaural cues~\cite{giurda2018reverberation},
and are thus essential for algorithms involving acoustic source localization, distance estimation,
spatial perception preservation, and speech intelligibility prediction~\cite{mershon1989effects,giguere1993sound,xia2018effects}.
Another issue is that most benchmarks are based on well-known datasets, which makes it possible for some models to exploit leaks between training and evaluation data~\cite{zhou2023don}.

We present and release the first benchmark dataset for the evaluation of audio representations in the context of hearables, called \gls{dear}.
The dataset consists of 1,158 mono tracks of 30 seconds each, obtained by mixing proprietary monologue recordings into acoustic scenes from a licensed database.
The benchmark includes eight classification and regression tasks covering general context, speech sources, and most importantly acoustic properties of audio scenes.
We compare four public self-supervised audio representation models and find that BEATs significantly outperforms all others.
In particular, we find that this model encodes technical acoustic properties considerably better than competitors, which gives important clues on how to design general-purpose audio models for hearables in the future.
The \gls{dear} dataset and associated code are available at \url{https://dear-dataset.github.io}.

\section{Methodology}
\label{sec:methodology}

\subsection{Dataset}
\label{ssec:dataset}

The \gls{dear} benchmark is generated by adding speech signals
to background sound scenes to ensure full control over the acoustic properties of the final mixture.
The background recordings were selected from the HOA-SSR dataset sound scene library (Force Technology, Denmark),%
\footnote{\url{https://forcetechnology.com/en/services/acoustics-noise-sound-quality/senselab-download-hoa-ssr-dataset}, retrieved September 9th, 2024.}
which is a curated collection of 150 audiovisual scenes captured using specialized equipment,
designed for comprehensive evaluations in audio product development.
In particular, we use the 4th-order ambisonics audio, which was recorded using an Eigenmike em32 and encoded in 25-channel AmbiX format at 48 kHz with a bit depth of 24.
The category selection has the purpose of capturing typical everyday situations.
The speech signals are proprietary anechoic monologues recorded with Lavalier microphones.
They span different vocal effort levels, which are elicited by playing pink noise through headphones at different levels.
The anechoic speech signals are then convolved with a set of impulse responses to produce sound mixtures with different combinations of speakers, positions, reverberation, and \glspl{snr}.
Throughout the process, attention was paid to avoid violations of the overall consistency of the generated sound scenes.

This procedure resulted in 1,158 audio tracks of 30s length with up to three speech signals, which may be active only at certain times.
Since the dataset is meant to explore public models that are mostly trained with mono-channel audio, the scenes are down-mixed to a single reference track at 44.1 kHz sample rate and bit depth of 32.
Each track is assigned to either model development or test to ensure fair comparison when using the dataset as a benchmark.
These two splits are generated using disjoint sets for all three components---monologues, background environment recordings, and impulse responses.
Monologues are all from distinct speakers, which prevents leakage, and each impulse response is associated with a specific background environment recording.

\subsection{Tasks}
\label{ssec:tasks}

Controlled dataset generation enables accurate labels at the track level, for each speech signal, and running annotations for each second.
Each track is assigned a type of environment, as well as a categorization as indoor or outdoor recording.
The reverberation of each speech signal as a whole is characterized by its \gls{drr} and its \gls{rt60}.
Furthermore, for each non-overlapping 1s segment, each speech signal is specified as active or not active, and in the active case the corresponding \gls{snr} is provided.
Finally, a dedicated set of tracks without speech signals contains only sound scenes dominated by noise, categorized as stationary or transient.

\begin{table}[t]
    \caption{
        List of downstream tasks used for the \acrshort{dear} evaluation, divided into three groups according to the audio properties they exploit, namely context, sources, and acoustics.
        Tasks are further categorized into binary or multi-class classification and regression, and as prospective or retrospective.
    }
    \label{tab:tasks}
    \centering
    \begin{tabular}{l|lll}
        \toprule
        \textbf{Task Name} & \textbf{Group} & \textbf{Objective} & \textbf{Type} \\
        \midrule
        Environment & \multirow{3}*{Context} & Multi-class & \multirow{3}*{Prospective} \\
        Indoor / Outdoor & & Binary & \\
        Stationary / Transient & & Binary & \\
        \midrule
        Speech Present & \multirow{2}*{Sources} & Binary & \multirow{2}*{Prospective} \\
        Speaker Count & & Regression & \\
        \midrule
        \acrshort{drr} & \multirow{3}*{Acoustics} & Regression & \multirow{3}*{Prospective} \\
        \acrshort{rt60} & & Regression & \\
        \acrshort{snr} & & Regression & \\
        \midrule
        TUT2017 & Context & Multi-class &  \multirow{2}*{Retrospective} \\
        LibriCount & Sources & Regression & \\
        \bottomrule
    \end{tabular}
\end{table}

\begin{figure*}[t]
    \centering
    \includegraphics[width=0.9\linewidth]{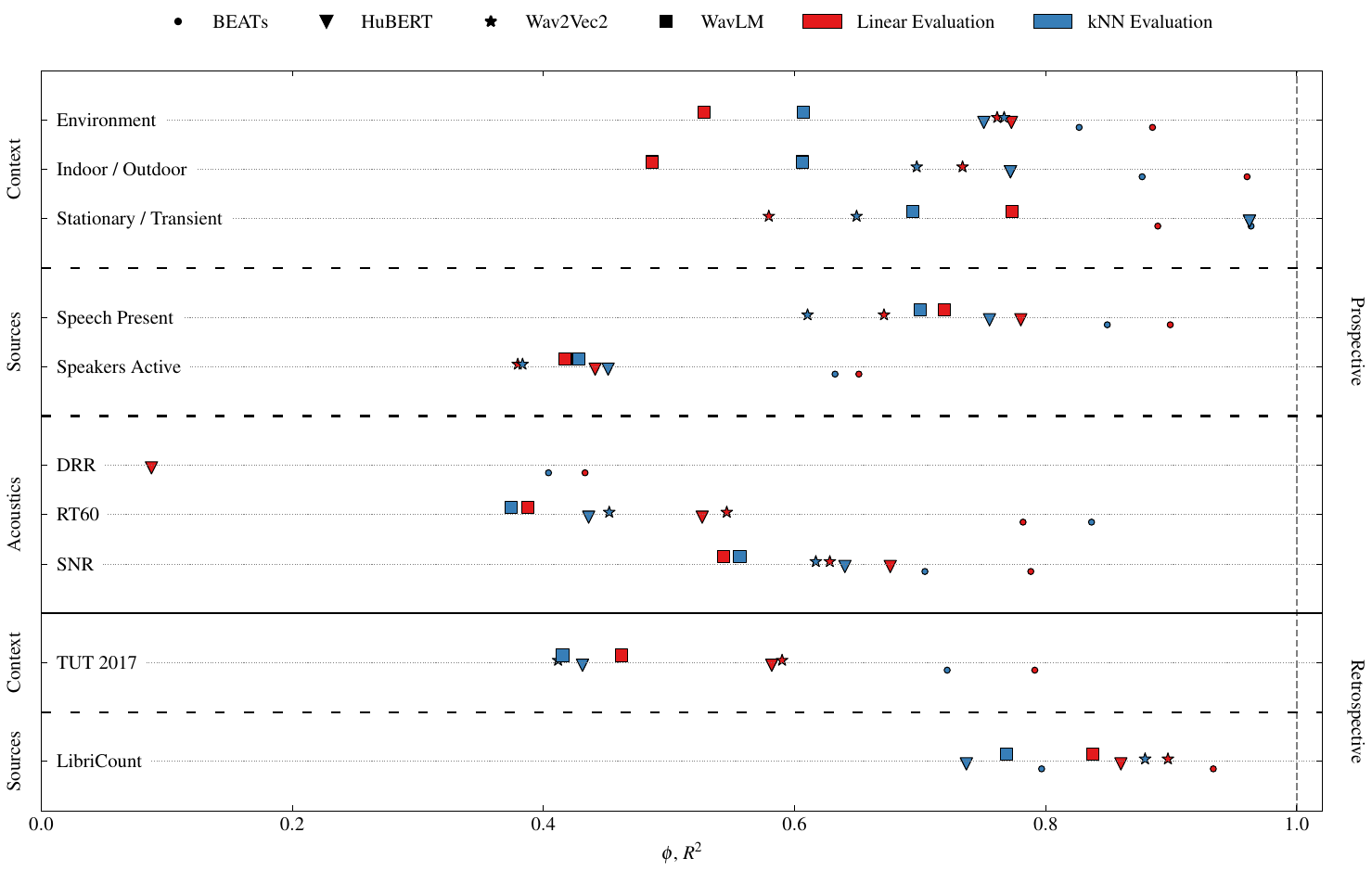}
    \caption{
        Comparison of the general-purpose audio representation models on a suite of downstream tasks.
        Evaluation is performed in terms of \acrshort{knn} and linear performance for three different task groups, namely context, sources, and acoustic properties.
        Regression tasks are evaluated in terms of $R^2$ and classification tasks in terms of Matthews' correlation coefficient $\phi$,
        markers missing from the plot have worse-than-random performance.
    }
    \label{fig:results}
\end{figure*}

The suite of evaluation tasks on the \gls{dear} dataset are summarized in the first three horizontal sections of \cref{tab:tasks} and divided into three groups.
Each group probes the ability of models to encode certain information, namely acoustic scene context for the first group, speech sources in the foreground for the second group, and acoustic properties of those sources for the third group.
For the acoustic scene context, we consider multi-class environment classification of whole tracks into one of five semantic categories: domestic, leisure, nature, professional, and transport, as well as binary classification of whole tracks as indoor or outdoor.
We also investigate binary noise classification as either stationary or transient using the corresponding tracks.
In this case we select the 30 seconds segment with the highest power to ensure sufficient information content.
To probe speech detection and speaker count in a relatively challenging setting,
we start from a set of tracks that contain at least one speech source,
and based on non-overlapping 1s segments, we classify whether speech is present or not, and regress the number of active speakers.
On the segments that have only one speech source active, we regress \gls{drr} in dB and \gls{rt60} on logarithmic scale, and on segments with only one speech source active, \gls{snr} in dB.

Given the commercial nature of the HOA-SSR dataset sound scene library including impulse responses, and the proprietary speech signals, the public models can be assumed to have never been trained or evaluated on the \gls{dear} data.
For this reason, we categorize the first eight tasks in \cref{tab:tasks}, which are the core of the benchmark, as prospective.
However, to establish a comparison with existing works, we also consider classification of acoustic scenes in one of 15 categories on the TUT2017 dataset~\cite{mesaros2016tut},
and regression of the number of speakers from 0 through 10 on LibriCount~\cite{stoter2018libricount},
with segments defined as in their original works.
Since it is possible that training and evaluation of the audio representation models considered in this work have already run on the same data used by these two tasks, we label them as retrospective.

\subsection{Models}
\label{ssec:models}

We selected publicly available general-purpose audio representation models, with the intent to cover a wide variety of different pre-training methods, and taking into account performance on typical representation benchmark datasets such as LibriSpeech~\cite{panayotov2015librispeech}, \mbox{AudioSet}~\cite{gemmeke2017audio}, and \mbox{ESC-50}~\cite{piczak2015esc}.
This resulted in the selection four models: Wav2Vec2.0~\cite{baevski2020wav2vec}, HuBERT~\cite{hsu2021hubert}, WavLM~\cite{chen2022wavlm}, and BEATs~\cite{chen2022beats}.
In this section, we first introduce the learning techniques underlying the four models.
We then discuss their common base architecture and specify their pre-training datasets.

Wav2Vec2.0~\cite{baevski2020wav2vec} improves upon the original Wav2Vec~\cite{schneider2019wav2vec} using a more advanced architecture including a convolutional neural network for extracting latent speech features, followed by a transformer network to capture contextual dependencies. 
The model 
uses a contrastive learning objective during pre-training, where it learns to distinguish true speech representations from distractors. 
HuBERT~\cite{hsu2021hubert} builds on the concept of masked prediction, similar to BERT~\cite{devlin2018bert} in natural language processing, adapted to speech. 
HuBERT works by learning to predict masked segments of continuous speech data using hidden units derived from clustering the speech features. 
The model is trained in a two-step process: first, it clusters speech features from a pre-trained model to generate pseudo-labels, and then it predicts these labels for masked portions of the input audio during training. 
WavLM~\cite{chen2022wavlm} builds upon the architecture of previous models like Wav2Vec2.0 and HuBERT. 
The model employs a transformer-based architecture and is trained using a masked speech prediction approach, where parts of the input audio are masked, and the model learns to predict these masked segments. 
BEATs~\cite{chen2022beats} uses acoustic tokenizers to convert continuous audio signals into discrete tokens, similar to how words are tokenized in text processing. 
The model uses a transformer-based architecture and learns to predict these discrete tokens from masked input audio segments during training, following a masked language model approach similar to WavLM and HuBERT. 
Pre-training is done in multiple iterations, and here we use the last iteration.

The architecture underlying all four models is a transformer of size ``base'' with 12 encoder layers, 768-dimensional hidden states, and 8 attention heads, resulting in 90 million parameters.
The publicly available models were pre-trained on different datasets, namely Wav2Vec2.0 and HuBERT on Librispeech~\cite{panayotov2015librispeech}, WavLM on LibriSpeech~\cite{panayotov2015librispeech}, GigaSpeech~\cite{chen2021gigaspeech}, and VoxPopuli~\cite{wang2021voxpopuli}, and BEATs on AudioSet~\cite{gemmeke2017audio}. 

\subsection{Evaluation}
\label{ssec:evaluation}

Following standard practice in representation learning, models are evaluated on unseen downstream tasks~\cite{turian2022hear}.
Specifically, the pre-trained models are used to obtain representations for audio segments, and a linear or \gls{knn} model is trained to solve each task based on frozen representations only.
Note that nearest neighbor evaluation probes the local structure of the feature space and has been shown to correlate well with fine-tuning performance \cite{caron2021emerging}.
This frozen evaluation examines how well representations encode a certain dataset attribute, indicating whether inherent properties of the downstream task were learned during pre-training.
The tracks reserved for evaluation in the \gls{dear} dataset are held out and used to compute the final performance, and the ones designated for development are further split into non-overlapping training and validation splits.
For \gls{knn}, the validation split is used to choose the number of neighbors among the values 1, 2, 5, 10, 20, and 50---separately for each model and task.
To ease result interpretation, we evaluate both binary and multi-class classification performance using Matthews' correlation coefficient $\phi$ and regression performance using the coefficient of determination $R^2$ over all testing tracks or 1s segments.
Both $\phi$ and $R^2$ take the value of 1 for perfect performance, and they are 0 for random models.

\section{Results}
\label{sec:results}

Figure \ref{fig:results} shows the performance of the four public pre-trained representation models described in section \ref{ssec:models} on the suite of downstream tasks presented in section \ref{ssec:tasks}, in terms of both \gls{knn} and linear performance.
Results indicate that BEATs outperforms other models across all tasks, often by a large margin.
Remarkably, this also holds in the group related to speech sources, even if the other three models were focused on speech analysis.
Although performance on the estimation of acoustic properties has room for improvement, BEATs obtains remarkable results, especially compared to its competitors.
This suggests that BEATs has successfully learned to encode \gls{snr} and some reverberation characteristics without explicit guidance towards these goals.
The success of the model could be attributed both to the label prediction objective from acoustic tokenizers as opposed to a contrastive objective and to the large and varied audio collection it was trained on, namely AudioSet.
A recent work suggests that the latter might be particularly important to capture multiple aspects of audio analysis at the same time~\cite{la2024benchmarking}.
Nevertheless, we observe that some tasks such as \gls{drr} estimation remain difficult and deserve separate investigation.

In the context group, which includes one 5-way and two binary classification tasks, all models demonstrate very good performance, even though BEATs still shows an advantage compared to the others.
Further insight can be obtained by looking at the public TUT2017 acoustic scene classification task, where model performance is lower.
This suggests that fine-grained classification is more difficult, which is plausible.
While all models encode the presence of speech, counting the number of active speakers proves to be challenging in the \gls{dear} dataset.
Looking at a similar publicly available task, LibriCount, reveals that models perform significantly better there than on the new benchmark.
This could be due to intrinsically more difficult noise backgrounds in \gls{dear}, or to the overlap between the pre-training and downstream data at least for HuBERT, Wav2Vec2.0, and WavLM. 

For LibriCount, TUT2017, and detecting the presence of speech, the performance under linear evaluation is typically slightly higher than its \gls{knn} counterpart.
This behavior can not be observed for the other tasks, and there is no clear pattern among the different evaluation types.

\section{Conclusion}
In this study, we introduced \gls{dear}, the first dataset and benchmark specifically designed to evaluate the performance of audio foundation models in encoding essential properties for effective steering in hearables---its general context, the presence and identification of speech sources, and technical acoustic properties such as reverberation and \gls{snr}.
Our experiments show that out of four compared models, BEATs strongly outperforms all competitors across eight prospective and two retrospective tasks.
For reverberation properties, the difference can be as high as 30 percentage points, and BEATs with \gls{knn} or linear regression effectively enables a qualitatively higher level of performance.
This benchmark sets the stage for the development of foundation models that are applicable to or even specialized for hearables, which will in turn enhance their signal processing capabilities and the overall user experience.
Audio foundation models like BEATs are currently still too complex for edge devices, but they allow sharing large parts of the computation for multiple tasks, and further advancements may soon reduce the gap to integration in hearable products.

\section*{Acknowledgment}
We convey the acknowledgment to FORCE Technology, Bang \& Olufsen, Demant, GN Store Nord, Sonova, WSA, and industrial partners who created the 360 audio-visual datasets under the HOA-SSR joint project, and to XRHub Team for the great help in dealing with technicalities and field recording.

\bibliographystyle{IEEEtran}
\bibliography{IEEEabrv,scholar}

\end{document}